# Single-Molecule Magnets: Ligand Induced Core Distortion and Multiple Jahn-Teller Isomerism in $[Mn_{12}O_{12}(O_2CMe)_8(O_2PPh_2)_8(H_2O)_4]$

Colette Boskovic,[†] Maren Pink,[†] John C. Huffman,[†] David N. Hendrickson*,[‡] and George Christou*,[†]

*Department of Chemistry and the Molecular Structure Center, Indiana University, Bloomington, IN 47405-7102. Department of Chemistry-0358, University of California at San Diego, La Jolla, CA 92093-0358*

Single-molecule magnets (SMMs) are a recent discovery that promises access to the ultimate high-density memory storage device in which each bit of digital information is stored on a single molecule. Each independent molecule of a SMM possesses the ability to function as a magnetizable magnet below a critical temperature.[1] This is due to intrinsic intramolecular properties, that is, a large spin ground state, and a large and negative (easy axis type) magnetoanisotropy, rather than to intermolecular interactions and long-range ordering. The most well studied SMMs are the $[Mn_{12}O_{12}(O_2CR)_{16}(H_2O)_x]^{n-}$ (n = 0-2) ($8Mn^{III}4Mn^{IV}$ for n = 0) complexes.[2,3,4] Evidence for the SMM behavior comes from frequency-dependent out-of-phase ac susceptibility signals ($\chi_M''$) and magnetization hysteresis loops. In addition, oriented crystals of these species display steps in the hysteresis loops indicative of field-tuned quantum tunneling of the magnetization (QTM).[5] A substantial amount of effort has been dedicated to systematic variation of the carboxylate R group and solvate molecules of crystallization, which has recently led to the discovery of Jahn-Teller (JT) isomerism,[6] involving differing relative orientations of the JT elongation axes of the eight $Mn^{III}$ centers and resulting in significantly different magnetic behavior for the different JT isomers. This includes differing positions of the $\chi_M''$ peaks and differing magnetization hysteresis plots, both of these effects reflecting different rates of magnetization relaxation.

We report herein a major expansion of the $Mn_{12}$ SMM family by demonstrating for the first time that carboxylate ligands can be replaced by other organic ligands, in this case diphenylphosphinate groups, to give $[Mn_{12}O_{12}(O_2CMe)_8(O_2PPh_2)_8(H_2O)_4]$. In addition, this species is unique in displaying three distinct JT isomers, two of which co-crystallize in the same crystal.

The reaction of $[Mn_{12}O_{12}(O_2CMe)_{16}(H_2O)_4]\cdot 2MeCO_2H\cdot 4H_2O$ (**1**) with eight equivalents of $Ph_2PO_2H$ in MeCN leads to crystallization of

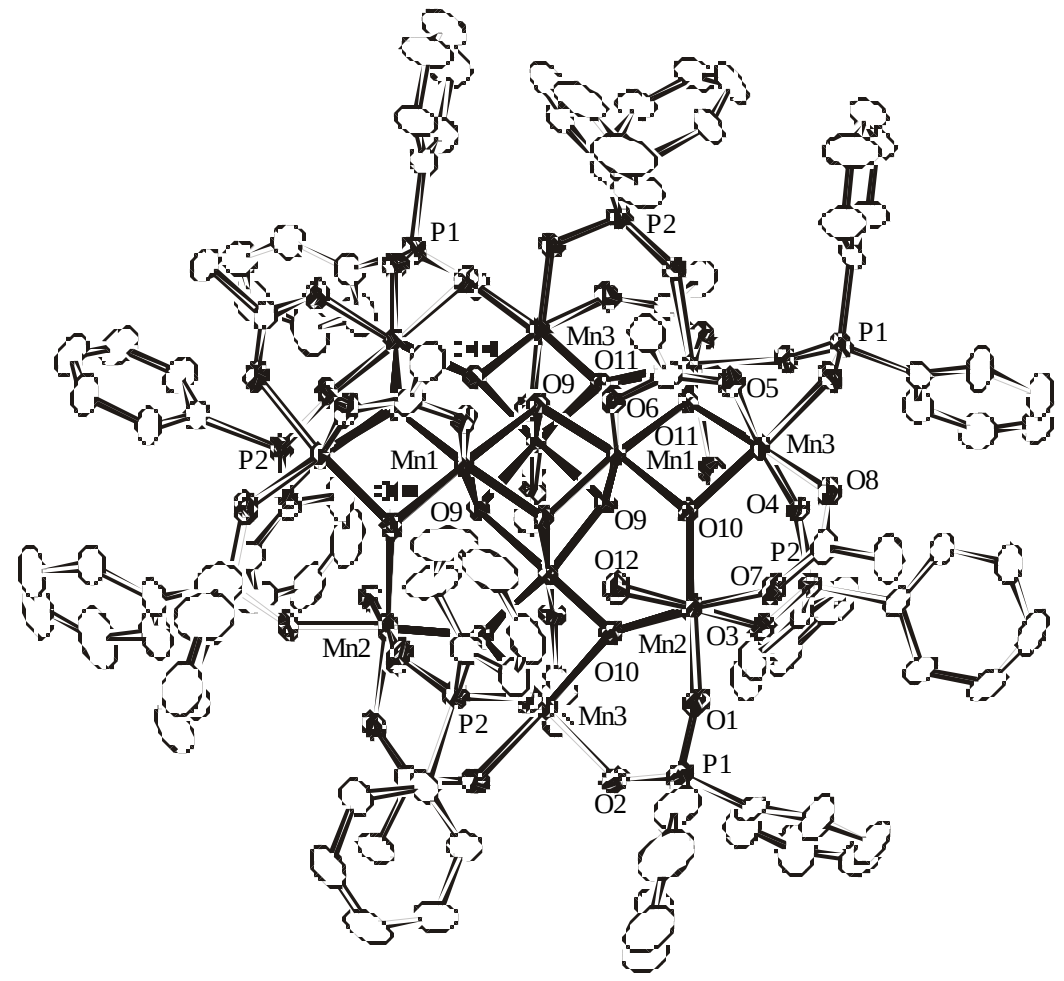

**Figure 1.** ORTEP representation of complex **2**·$12CH_2Cl_2$ at the 50% probability level.

$[Mn_{12}O_{12}(O_2CMe)_8(O_2PPh_2)_8(H_2O)_4]$ (**2**) in 60% yield, which was crystallographically characterized as **2**·$14/3MeCN\cdot 4/3H_2O$.[7] The material can be recrystallized from $CH_2Cl_2$/hexanes to give **2**·$12CH_2Cl_2$.[8] **2**·$14/3MeCN\cdot 4/3H_2O$ crystallizes in space group *C2/c* with two independent molecules in the structure, the asymmetric unit containing 1.5 $Mn_{12}$ clusters in addition to solvent. **2**·$12CH_2Cl_2$ crystallizes in space group $P4_2/n$, with the asymmetric unit containing 0.25 of the $Mn_{12}$ cluster and solvent. All three crystallographically unique molecules possess a structure similar to that of **1**, with a central $[Mn^{IV}_4O_4]$ cubane held within a non-planar ring of eight $Mn^{III}$ ions by eight $\mu_3$-$O^{2-}$ ions. However, in **2** the four axial $Mn^{III}$-$Mn^{III}$ and four of the eight equatorial $Mn^{III}$-$Mn^{III}$ carboxylate sites have been replaced by diphenylphosphinate, while the remaining equatorial sites and the four axial $Mn^{IV}$-$Mn^{III}$ sites remain as acetates (Figure 1), resulting in a significant distortion of the $[Mn_{12}O_{12}]^{16+}$ core (Figure 2). This is reflected in the bond angles, in an increase of ~ 0.1 Å in all of the $Mn^{III}$-$Mn^{III}$ distances in **2** and is most apparent as a "bowing" in each of the linearly arranged $Mn_4$ units, with the angles of the type Mn(3)-Mn(1)-Mn(1′) in **2**·$12CH_2Cl_2$, decreasing from ~ 178° in **1** to 171–173° in **2**. All three of the molecules possess one $H_2O$ ligand bound to every other $Mn^{III}$ center (Mn(2) for **2**·$12CH_2Cl_2$), oriented alternately in opposite directions, as is observed in **1**. The primary difference between the structures of the different molecules of **2** lies in the location of the JT elongation axes of the eight $Mn^{III}$ centers (Figure 2). The single cluster in **2**·$12CH_2Cl_2$ possesses approximate $S_4$ symmetry, with all eight JT axes oriented approximately parallel in the axial direction. In this case, none of the JT axes are directed towards the oxide ligands, as is the usual case in $Mn_{12}$-carboxylate species, including **1**. However, the two independent molecules in **2**·$14/3MeCN\cdot 4/3H_2O$ possess $C_1$ and approximate $S_4$ (crystallographic $C_2$) symmetry, with two (Mn(9) and Mn(11)) and four (Mn(185), Mn(185′), Mn(186) and Mn(186′))

† Indiana University.
‡ University of California at San Diego.

[7] Crystal data for 3/2[**2**]·$7MeCN\cdot 2H_2O$: $C_{182}H_{193}N_7Mn_{18}O_{74}P_{12}$, 5023.11 g $mol^{-1}$, monoclinic, *C2/c*, a = 55.071(4) Å, b = 33.491(3) Å, c = 27.331(2) Å, β = 114.966(2)°, Z = 8, V = 45699.0 $Å^3$, $d_{calc}$ = 1.460 $gcm^{-3}$, T = -158 °C. The structure was solved by direct methods (SHELXTL) and refined (on F) using 52187 reflections with F>3σ(F) to R($R_w$) values of 0.0777 (0.0645).

[8] Crystal data for **2**·$12CH_2Cl_2$: $C_{124}H_{136}Cl_{24}Mn_{12}O_{48}P_8$, 4152.17 g $mol^{-1}$, tetragonal, $P4_2/n$, a = b = 25.208(2) Å, c = 13.413(1) Å, Z = 2, V = 8523.1 $Å^3$, $d_{calc}$ = 1.618 $gcm^{-3}$, T = -160 °C. The structure was solved by direct methods (SHELXTL) and refined (on F) using 9776 reflections with F>2σ(F) to R($R_w$) values of 0.0666 (0.2180).

equatorially-oriented JT axes, respectively. That is, for these species, some of the JT

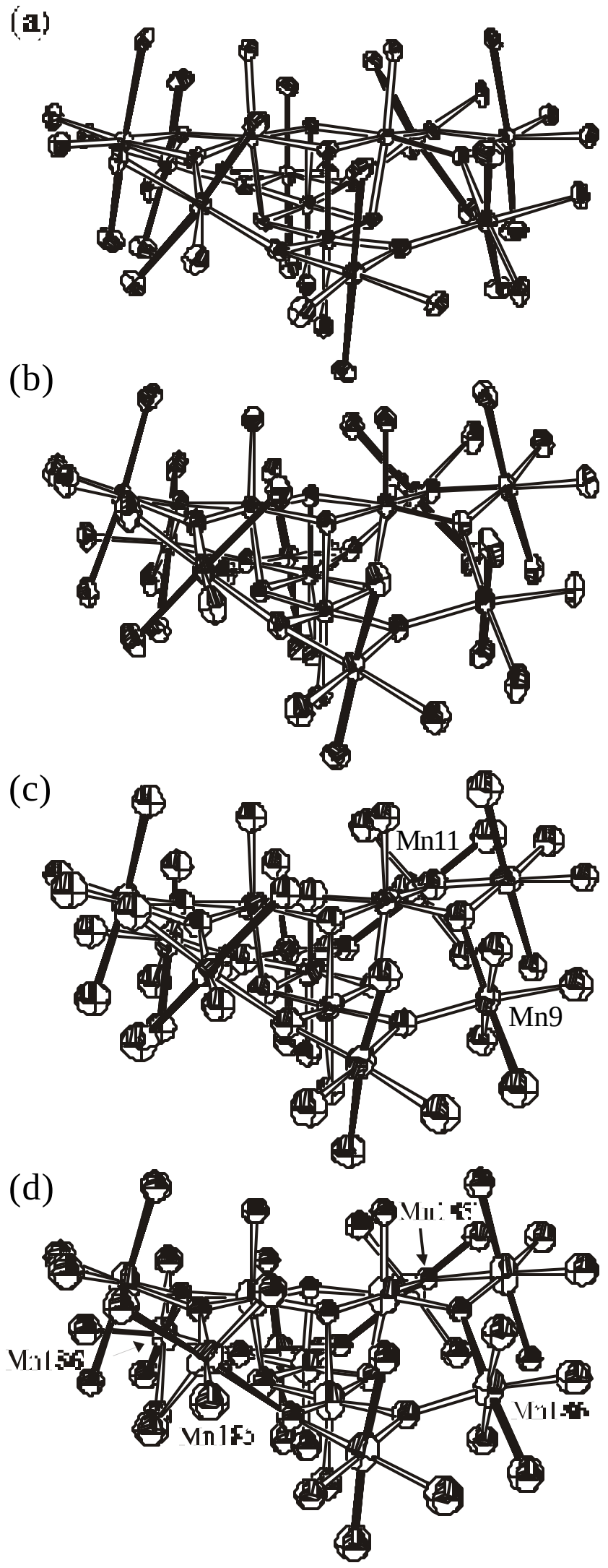

**Figure 2.** The cores of complexes: (a) **1**, (b) **2**·$12CH_2Cl_2$, (c) the $C_1$ molecule of **2**·14/3MeCN·4/3$H_2O$ and (d) the $S_4$ molecule of **2**·14/3MeCN·4/3$H_2O$, emphasizing the Jahn-Teller axes.

axes point towards oxide ligands. In both cases, the "abnormally" oriented JT axes occur at the $Mn^{III}$ centers with $H_2O$ ligands, as has been observed for the previously reported examples of such JT isomerism.[6]

The ground state of **2**·$12CH_2Cl_2$ was determined from reduced magnetization ($M/N\mu_B$) *vs.* H/T measurements in the 1.8-25 K and 20-70 kG range.[9] Fitting of the data[10] gave S = 10, g = 1.92 and D = -0.41 $cm^{-1}$ (-0.59 K), similar to the values generally observed for $[Mn_{12}O_{12}(O_2CR)_{16}(H_2O)_x]$ clusters (S = 10, D ~ -0.5 $cm^{-1}$).

AC magnetization measurements were performed on **2**·$12CH_2Cl_2$ in the 1.8-10 K range in a 3.5 G AC field oscillating at 1-1500 Hz. The in-phase $\chi_M'T$ signal shows a frequency-dependent decrease at T < 8 K, indicative of the onset of slow relaxation. This was confirmed by the concomitant appearance of an out-of-phase ($\chi_M''$) signal due to the inability of **2**·$12CH_2Cl_2$ to relax sufficiently rapidly at these temperatures to keep up with the oscillating field. For the ac frequency range 1-1500 Hz, the $\chi_M''$ peaks of

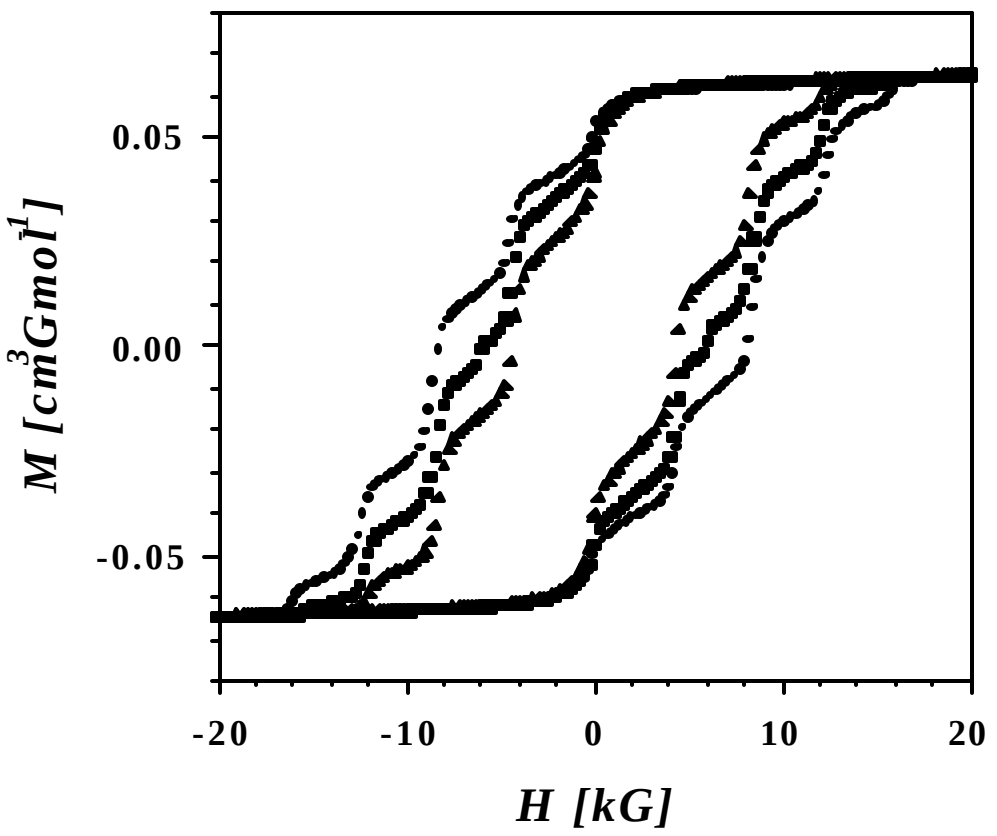

**Figure 3.** Magnetization hysteresis loops for oriented crystals measured in an eicosane matrix for complex **2**·$12CH_2Cl_2$ at 1.8 (•), 1.9 (■) and 2.0 K (▲).

**2**·$12CH_2Cl_2$ occur in the range 3-7 K, again consistent with the behavior generally observed for $[Mn_{12}O_{12}(O_2CR)_{16}(H_2O)_x]$. Data obtained by varying the frequency of oscillation of the AC field can be fit to an Arrhenius equation to give the effective energy barrier for the relaxation of the magnetization ($U_{eff}$).[9] An analysis of the variable frequency $\chi_M''$ data indicates that $U_{eff}$ for complex **2**·$12CH_2Cl_2$ is ~ 42 $cm^{-1}$ (60 K), which again is within the range normally observed for $[Mn_{12}O_{12}(O_2CR)_{16}(H_2O)_x]$ of 42-50 $cm^{-1}$ (60-72 K). AC measurements on **2**·14/3MeCN·4/3$H_2O$ are in progress, but are hindered by crystal degradation following rapid solvent loss.

Magnetization hysteresis loops were observed for oriented crystals of **2**·$12CH_2Cl_2$ at 1.8, 1.9 and 2.0 K (Figure 3). A few crystals were suspended in eicosane at 40 °C, oriented in a 70 kG field to allow the crystals to align their principal axis of magnetization parallel to the applied field, and then cooled to room temperature. A number of steps are evident in the hysteresis loop, due to quantum tunneling of the magnetization. The separation between steps was found to be 3.95 kG. This indicates a value of D/g of ~ -0.18 $cm^{-1}$, which is consistent with the value of -0.21 $cm^{-1}$ obtained from the fitting of the reduced magnetization.

In conclusion, the first significantly altered derivative of $[Mn_{12}O_{12}(O_2CR)_{16}(H_2O)_x]$ has been prepared by incorporation of non-carboxylate organic ligands, and ligand-induced core distortions result. In addition, multiple JT isomerism has been observed, emphasizing the small energy differences involved. Complex **2** is magnetochemically similar to its 16-carboxylate parent, possessing an S = 10 ground state and displaying frequency-dependent peaks in the out-of-phase AC susceptibility plots in addition to magnetization hysteresis. Furthermore, quantum tunneling of the magnetization is evident as steps in the hysteresis loops. Thus the SMM properties are still retained in the diphenylphosphinate-substituted species **2**, which thus represents the prototype of a major new thrust in the SMM field.

**ACKNOWLEDGMENT.** This work was supported by National Science Foundation grants to G.C. and D.N.H.

---

[9] See Supporting Information.
[10] Vincent, J. B.; Christmas, C.; Chang, H.-R.; Li, Q.; Boyd, P. D. W.; Huffman, J. C.; Hendrickson, D. N.; Christou, G. *J. Am. Chem. Soc.* **1989**, *111*, 2086.

**SUPPORTING INFORMATION AVAILABLE.** Crystallographic details for **2**·12$CH_2Cl_2$ and **2**·14/3MeCN·4/3$H_2O$ (PDF). Reduced magnetization plot and Arrhenius plot. This material is available free of charge via the Internet at http://pubs.acs.org.